\documentclass[twocolumn]{article}
\usepackage{amssymb}

\usepackage{amsmath}

\newtheorem{theorem}{Theorem}
\newtheorem{acknowledgement}[theorem]{Acknowledgement}

\input{tcilatex}

\begin{document}

\title{Short-time decoherence of Josephson charge qubits in Ohmic and $1/f$
noise environment}
\author{Xian-Ting Liang\thanks{%
Email address: xtliang@ustc.edu} \\
Department of Physics and Institute of Modern Physics,\\
Ningbo University, Ningbo 315211, China}
\maketitle

\begin{abstract}
In this paper we investigate the short-time decoherence from Ohmic and $1/f$
noise of single Josephson charge qubit (JCQ). At first, we use the
short-time approximation to obtain the dynamics of the open JCQ. Then we
calculate the decoherence the measure of which is chosen as the maximum norm
of the deviation density operator. It is shown that the decoherence from $%
1/f $ noise plays the central role. The total decoherence from Ohmic and $%
1/f $ noise is serious at present experiential conditions according to the
DiVincenzo criterion.


\end{abstract}

\section{Introduction}

Quantum bit (qubit) is a key block for building quantum computers. Among
various realizations of the qubit for quantum computations that based on the
Josephson junctions are considered to be particularly promising candidates
because of their scalability, established fabrication techniques, and
flexibility in designs. Many kinds of the superconducting
Josephson-junction-qubit models are proposed in last years. They are
Josephson charge \cite{charge-qubit}, flux \cite{flux-qubit}, phase \cite%
{phase-qubit} and hybridized \cite{hybrid-qubit} qubits. According to
DiVincenzo \cite{DiVincenzo} proposal, the qubit for quantum computation
must satisfy five criteria one of which is the low decoherence criterion. An
approximate benchmark of the criterion is a fidelity loss no more than $\sim
10^{-4}$ per elementary quantum gate operation. The decoherence of the
Josephson single qubit and coupling qubits has been investigated widely in
last years \cite{decoherence-JCQ}. In these researches the dynamics of the
qubits interacting with their environment has been treated by the
perturbation method, the path integral method \cite{perturbation}, and
others \cite{Gaussian}. In the researches in general some approximation
scheme must be used. The most familiar and frequently used one is the Markov
approximation \cite{Markov}.

It has been pointed that the Markov approximation can not be used in low
temperature and for short cycle times of quantum computation \cite{Markov
can not}. However, the qubits based on the Josephson superconducting
junction should be worked in low temperature, such as few 10 mK. So the
Markov approximation can not be used in the investigations of decoherence
for Josephson qubits. Fortunately, a short-time approximation scheme \cite%
{short-time}\emph{\ }being\emph{\ }fit for the investigations has been
developed recently by Privman \emph{et al. }\cite{Privman,Fedichkin et al}.
The approaches of V. Privman and his co-worker are rather formal and
universal. It is also very interesting to investigate the decoherence of a
concrete and physical qubit model because it can help us to know that
whether the model satisfy the DiVincenzo low decoherence criterion or not.
If the model can not satisfy the criterion, the qubit has then not
qualification for being a quantum computation hardware.

In this paper we shall investigate that if the JCQ model satisfy the
DiVincenzo low decoherence criterion. The same problem has been investigated
in our previous works \cite{Liang01,Liang02} where we suppose that there is
only the Ohmic noise in the environment of JCQ. However, recent several
experiments ( see \cite{1/f-noise-JCQ} and within ) with Josephson junction
circuits have revealed at low frequencies the presence of $1/f$ noise and it
is shown that the $1/f$ noise plays a major role in destructing the
coherence of qubit in solid-state systems. In this paper, we shall carefully
estimate the decoherence of the JCQ in Ohmic and $1/f$ noise environment.
Here, the measure of decoherence is based on a standard operator norm $%
\left\| A\right\| $ in the theory of linear operators \cite{sigma}. The
short-time approximation of the split-operator will be used in our
investigation.

\section{Open JCQ model}

The single JCQ Hamiltonian is \cite{PRL1997,RMP2001} 
\begin{equation}
H_{s}=4E_{c}\left( n-n_{g}\right) ^{2}-E_{J}\cos \varphi .  \label{eq1}
\end{equation}%
Here, $E_{c}=e^{2}/2\left( C_{g}+C_{J}\right) $ is the charging energy; $%
E_{J}=I_{c}\hbar /2e$ is the Josephson coupling energy \cite{SM25-915-1999},
where $I_{c}$ is the critical current of the Josephson junction, $\hbar $
the Planck's constant divided $2\pi ,$ and $e$ the charge of electron; $%
n_{g}=C_{g}V_{g}/2e=Q_{g}/2e$ is the dimensionless gate charge, where $C_{g}$
is the gate capacitance, $V_{g}$ the controllable gate voltage. The number
operator $n$ of (excess) Cooper pair on the island, and the phase operator $%
\varphi $ of the superconducting order parameter, are quantum mechanically
conjugate \cite{RMP2001}. The unavoidable noise of the environment may lead
to the dissipation. In general, one takes into account that the environment
itself is also a quantum system with a large number degrees of freedom.
Usually it is modeled by a bath with a large set of harmonic oscillators %
\cite{Weissbook,RMP59-1-1987}%
\begin{equation}
H_{B}=\tsum\nolimits_{\alpha }\left[ \frac{1}{2m_{\alpha }}p_{\alpha }^{2}+%
\frac{1}{2}m_{\alpha }\omega _{\alpha }^{2}x_{\alpha }^{2}\right] ,
\label{eq2}
\end{equation}%
each of which interacts weakly with the system of interest. The whole
Hamiltonian of the system-bath is \cite{open-JCQ-model} then 
\begin{eqnarray}
H &=&H_{s}+\tsum\nolimits_{\alpha }\left[ \frac{1}{2m_{\alpha }}p_{\alpha
}^{2}\right.  \notag \\
&&\left. +\frac{1}{2}m_{\alpha }\omega _{\alpha }^{2}\left( x_{\alpha }-%
\frac{\lambda _{\alpha }}{m_{\alpha }\omega _{\alpha }^{2}}2en\frac{C_{t}}{%
C_{J}}\right) ^{2}\right] .  \label{eq3}
\end{eqnarray}%
Here, the coupling operator of the environment is $X=\tsum \lambda _{\alpha
}x_{\alpha }.$ If the Josephson coupling energy $E_{J}$ is much smaller than
the charging energy $E_{c}$, and both of them are much smaller than the
superconducting energy gap $\Delta $, the Hamiltonian $H_{s}$ of Josephson
junction can be parameterized by the number of the Cooper pairs $n$ on the
island. When the temperature $T$ is low enough the system can be reduced to
a two-state system (qubit) because all other charge states have much higher
energy and can be neglected. So the Hamiltonian of the system can
approximately reads $H_{s}=-\frac{1}{2}B_{z}\sigma _{z}-\frac{1}{2}%
B_{x}\sigma _{x},$ where $B_{z}=E_{ch}\left( 1-2n_{g}\right) $ and $%
B_{x}=E_{J}.$ If we choose the working point to make $n_{g}$ half-integer,
say $n_{g}=1/2,$ we can obtain that%
\begin{equation}
H\equiv H_{s}+H_{I}+H_{B}.  \label{eq4}
\end{equation}%
Here,%
\begin{eqnarray}
H_{s} &=&-\frac{1}{2}B_{x}\sigma _{x},\text{ }H_{B}=\tsum\nolimits_{\alpha
}\hbar \omega _{\alpha }a_{\alpha }^{\dagger }a_{\alpha },  \notag \\
H_{I} &=&-\sigma _{z}\sqrt{\hbar }\tsum g_{\alpha }\left( a_{\alpha
}^{\dagger }+a_{\alpha }\right) ,  \label{eq5}
\end{eqnarray}%
where%
\begin{equation}
g_{\alpha }=\lambda _{\alpha }\sqrt{\frac{1}{m_{\alpha }\omega _{\alpha }}}%
\left( 2e\frac{C_{t}}{C_{J}}\right) .  \label{eq6}
\end{equation}

\section{Ohmic and $1/f$ noise}

Recent studies, both the experimental and theoretical, suggest that the
serious noise sources in Josephson devices include two kinds of fluctuations
one of which with ohmic spectrum the other with $1/f$ spectrum \cite%
{1/f-noise-JCQ}. In the following, we introduce the spectral densities of
the Ohmic and $1/f$ noise. According to \cite{Weissbook}, from Eq. (\ref{eq3}%
), we can express the spectral density of the noise as%
\begin{align}
J_{X}\left( \omega \right) & =\frac{\pi }{2}\tsum_{\alpha }\left( 2e\frac{%
C_{t}}{C_{J}}\right) ^{2}\frac{\left| \lambda _{\alpha }\right| ^{2}}{%
m_{\alpha }\omega _{\alpha }}\delta \left( \omega -\omega _{\alpha }\right) 
\notag \\
& =\frac{\pi }{2}\hbar \tsum_{\alpha }\left| g_{\alpha }\right| ^{2}\delta
\left( \omega -\omega _{\alpha }\right)  \notag \\
& =\frac{\pi }{2}\hbar \int_{-\infty }^{\infty }\left| g\left( \omega
_{\alpha }\right) \right| ^{2}\delta \left( \omega -\omega _{\alpha }\right)
\notag \\
& =\frac{\pi }{2}\hbar D\left( \omega \right) \left| g\left( \omega \right)
\right| ^{2},  \label{eq7}
\end{align}%
where $D\left( \omega \right) $ is the density of the states of environment
modes. On the other hand, the spectral density is related to the power
spectrum as \cite{Weissbook,open-JCQ-model}%
\begin{equation}
S_{X}\left( \omega \right) =J_{X}\left( \omega \right) \hbar \coth \left(
\omega \beta /2\right) .  \label{eq8}
\end{equation}%
Here, $\beta =\hbar /k_{B}T,$ where $k_{B}$ is the Boltzmann constant. It
has been stressed that the relationship (\ref{eq8}) are satisfied not only
for the linear coupling model but also for the nonlinear one. To the Ohmic
noise case, from the fluctuation-dissipation theorem one has \cite%
{open-JCQ-model} 
\begin{align}
\left. S_{X}\left( \omega \right) \right| _{O}& =\left( \frac{2eC_{t}}{C_{J}}%
\right) ^{2}\left. S_{V}\left( \omega \right) \right| _{O}  \notag \\
& =\left( \frac{2eC_{t}}{C_{J}}\right) ^{2}\func{Re}Z_{t}\left( \omega
\right) \omega \hbar \coth \left( \omega \beta /2\right) .  \label{eq9}
\end{align}%
Then, setting $\func{Re}Z_{t}\left( \omega \right) \approx R,$ one has%
\begin{equation}
\left. J_{X}\left( \omega \right) \right| _{O}=\left( 2e\right) ^{2}\left( 
\frac{C_{t}}{C_{J}}\right) ^{2}R\omega .  \label{eq10}
\end{equation}%
So by using Eq. (\ref{eq7}) we have%
\begin{equation}
\left. D\left( \omega \right) g^{2}\left( \omega \right) \right| _{O}=4\frac{%
\left( 2e\right) ^{2}}{\hbar }\left( \frac{C_{t}}{C_{J}}\right) ^{2}R\omega .
\label{eq11}
\end{equation}%
Considering the cutoff of the frequency $\omega $ we can set%
\begin{equation}
\left. D\left( \omega \right) g^{2}\left( \omega \right) \right| _{O}=\eta
\omega \exp \left( \frac{\omega }{\omega _{c}}\right) .  \label{eq12}
\end{equation}%
Here, $\omega _{c}$ is the cutoff frequency and $\eta =4\frac{R}{R_{Q}}%
\left( \frac{C_{t}}{C_{J}}\right) ^{2}$, where $R_{Q}=h/\left( 2e\right)
^{2}\approx 6.5$ k$\Omega $ \cite{note}$.$ In the model of the JCQ circuit
the typical impedance of the control line is $R\approx 50$ $\Omega ,$ and $%
C_{g}\approx 10^{-18}$ F$,$ $C_{J}\approx 10^{-16}$ F$,$ so we can obtain $%
\eta \approx 10^{-6}$.

The $1/f$ noise is considered deriving from the background charge
fluctuations in the circuits. It can be expressed an effective noise of the
gate charge, i. e., $S_{Q_{g}}\left( \omega \right) =\alpha _{f}e^{2}/\omega 
$. Recent experiments proposed at relevant temperatures $\alpha _{f}\sim
10^{-7}-10^{-6}.$ According to Ref. \cite{open-JCQ-model} this noise can be
translated into the fluctuations of $X$, namely,%
\begin{equation}
\left. S_{X}\left( \omega \right) \right| _{f}=E_{f}^{2}/\omega ,
\label{eq13}
\end{equation}%
where $E_{f}=4E_{c}\sqrt{\alpha _{f}}.$ Comparing the Eq. (\ref{eq13}) with
Eq. (\ref{eq8}) we have 
\begin{equation}
\left. J_{X}\left( \omega \right) \right| _{f}=\frac{16E_{c}^{2}\alpha _{f}}{%
\hbar \omega \coth \left( \omega \beta /2\right) }.  \label{eq14}
\end{equation}%
So%
\begin{align}
\left. D\left( \omega \right) g^{2}\left( \omega \right) \right| _{f}& =%
\frac{32E_{c}^{2}\alpha _{f}}{\pi \hbar ^{2}\omega \coth \left( \omega \beta
/2\right) }  \notag \\
& =\frac{\kappa \alpha _{f}}{\omega \coth \left( \omega \beta /2\right) },
\label{eq15}
\end{align}%
where 
\begin{equation}
\kappa =\frac{64E_{c}^{2}}{h\hbar }\approx 1.5\times 10^{25}.  \label{eq16}
\end{equation}

In fact, in a JCQ circuit these two kinds of noise are existed at the same
time. So the spectral density of the total noise is%
\begin{equation}
J\left( \omega \right) =\left. J_{X}\left( \omega \right) \right|
_{O}+\left. J_{X}\left( \omega \right) \right| _{f}.  \label{eq17}
\end{equation}%
So we have 
\begin{equation}
D\left( \omega \right) g^{2}\left( \omega \right) =\left. D\left( \omega
\right) g^{2}\left( \omega \right) \right| _{O}+\left. D\left( \omega
\right) g^{2}\left( \omega \right) \right| _{f}.  \label{eq18}
\end{equation}%
In the following, we shall use above three kinds of different noise of the
environment investigating the decoherence of the JCQ.

\section{Short-time dynamics of JCQ}

Before studying the decoherence we shall investigate the dynamics of the
open JCQ with the short-time approximation. Suppose the initial state of the
JCQ-bath be $R\left( 0\right) =\rho \left( 0\right) \otimes \Theta ,$ where $%
\rho \left( 0\right) $ is the initial state of JCQ and $\Theta $ is the
initial state of the environment. We set $\Theta $ is the product of the
bath modes density matrices\ $\theta _{k}$. In the initial states, each bath
mode $k$ is assumed to be thermalized, namely,%
\begin{equation}
\theta _{k}=\frac{e^{-\beta M_{k}}}{\text{Tr}_{k}\left( e^{-\beta
M_{k}}\right) },  \label{eq19}
\end{equation}%
where $M_{k}=\omega _{k}a_{k}^{\dagger }a_{k}$. The evolution operator of
the JCQ-bath is then%
\begin{equation}
U=e^{-iH\tau /\hslash }=e^{-i\left( H_{s}+H_{I}+H_{B}\right) \tau /\hslash }.
\label{eq20}
\end{equation}%
Due to non-conservation of $H_{s}$ in this system, the evolution operator
cannot be in a general way expressed as $e^{-iH_{s}\tau /\hbar }e^{-i\left(
H_{I}+H_{B}\right) \tau /\hbar }.$ But in the sort-time approximation, the
operator can be approximately expressed as \cite%
{split-operator01,split-operator02}%
\begin{equation}
U=e^{-iH_{s}\tau /2\hbar }e^{-i\left( H_{I}+H_{B}\right) \tau /\hbar
}e^{-iH_{s}\tau /2\hbar }+o(\tau ^{3}).  \label{eq21}
\end{equation}%
It has been proved that the expression is accurate enough for the time being
short to the characteristic time. In the following we only investigate the
case that the system evolute within the time $\tau <10$ ps, the
characteristic time of the single JCQ is $\tau =12.7$ ps (when $E_{J}=51.8$ $%
\mu $eV) \cite{Liang02}. So the elements of the reduced density matrix $\rho
\left( \tau \right) $ in the basis of operator $H_{s}$ can be expressed as%
\begin{align}
\rho _{mn}& =\text{Tr}_{B}\left\langle \varphi _{m}\right| e^{-iH_{s}\tau
/2\hbar }e^{-i\left( H_{I}+H_{B}\right) \tau /\hbar }e^{-iH_{s}\tau /2\hbar
}R\left( 0\right)  \notag \\
& e^{iH_{s}\tau /2\hbar }e^{i\left( H_{I}+H_{B}\right) \tau /\hbar
}e^{iH_{s}\tau /2\hbar }\left| \varphi _{n}\right\rangle ,  \label{eq22}
\end{align}%
where $\left\{ m,n\right\} =0$ or $1$. The $\rho $ is a matrix with $2$ by $%
2 $. In the following we set $t=\tau /\hbar .$ Through some calculations we
can obtain the evolution of the density matrix elements $\rho _{10}\left(
t\right) $ and $\rho _{11}\left( t\right) $\ as \cite{Liang01,Liang02}%
\begin{align}
\rho _{10}\left( t\right) & =\frac{1}{2}\rho _{10}\left( 1-e^{-B^{2}\left(
t\right) }+e^{itE_{J}}+e^{itE_{J}-B^{2}\left( t\right) }\right) ,  \notag \\
\rho _{11}\left( t\right) & =\frac{1}{2}\rho _{00}\left( 1-e^{-B^{2}\left(
t\right) }\right) +\frac{1}{2}\rho _{11}\left( 1+e^{-B^{2}\left( t\right)
}\right) ,  \label{eq23}
\end{align}%
where $\rho _{00}=\rho _{00}\left( 0\right) ,$ $\rho _{11}=\rho _{11}\left(
0\right) $, $\rho _{10}=\rho _{10}\left( 0\right) ,$ and 
\begin{align}
B^{2}\left( t\right) & =8\tsum_{k}\frac{\left| g_{k}\right| ^{2}}{\omega
_{k}^{2}}\sin ^{2}\frac{\omega _{k}t}{2}\coth \frac{\beta \omega _{k}}{2}, 
\notag \\
C\left( t\right) & =\tsum_{k}\frac{\left| g_{k}\right| ^{2}}{\omega _{k}^{2}}%
\left( \omega _{k}t-\sin \omega _{k}t\right) .  \label{eq24}
\end{align}%
When the summation in Eq. (\ref{eq24}) is converted to integration in the
limit of infinite number of the bath modes, one has%
\begin{equation}
B^{2}\left( t\right) =8\int d\omega D\left( \omega \right) g\left( \omega
\right) ^{2}\omega ^{-2}\sin ^{2}\frac{\omega t}{2}\coth \frac{\beta \omega 
}{2},  \label{eq25}
\end{equation}%
for the real $g\left( \omega \right) .$ In fact this $B^{2}\left( t\right) $
is the mean-squared value of the magnitude of the phase noise in Ref. \cite%
{PRB67-094510}. From the deriving of $B^{2}\left( t\right) $ we know that it
results from the choice of bath model and the coupling forms of JCQ-bath
rather than the choice of the approximation scheme. But the evolution of the
matrix elements as Eqs. (\ref{eq23}) is resulted from the short-time
approximation and it can not be obtained by using the Markov approximation
and others. By using above results we can yield a good quantitative
estimation to the decoherence behavior of JCQ in a short time.

\section{Decoherence of JCQ}

It is shown that the measure 
\begin{equation}
D\left( t\right) =\sup_{\rho \left( 0\right) }\left( \left\| \sigma \left(
t,\rho \left( 0\right) \right) \right\| _{\lambda }\right)  \label{eq26}
\end{equation}%
is suitable for estimating the decoherence of the qubit gates \cite%
{Privman,Fedichkin et al}. Here, the norm $\left\| \sigma \right\| _{\lambda
}$ is defined as%
\begin{equation}
\left\| \sigma \right\| _{\lambda }=\sup_{\varphi \neq 0}\left( \frac{%
\left\langle \varphi \right| \sigma \left| \varphi \right\rangle }{%
\left\langle \varphi \right. \left| \varphi \right\rangle }\right) ^{\frac{1%
}{2}}.  \label{eq27}
\end{equation}%
For a qubit, it is%
\begin{equation}
\left\| \sigma \right\| _{\lambda }=\sqrt{\left| \sigma _{10}\right|
^{2}+\left| \sigma _{11}\right| ^{2}}.  \label{eq28}
\end{equation}%
Here, the deviation operator $\sigma $ is defined as 
\begin{equation}
\sigma \left( \tau \right) =\rho \left( \tau \right) -\rho ^{i}\left( \tau
\right) ,  \label{eq29}
\end{equation}%
where $\rho \left( \tau \right) $ and $\rho ^{i}\left( \tau \right) $ are
density matrixes of the ``real''\ evolution (with interaction) and the
``ideal''\ one (without interaction) of the investigated system. The
evolution of the closed JCQ is $\rho _{11}^{i}\left( t\right) =\rho _{11},$
and $\rho _{10}^{i}\left( t\right) =\rho _{10}e^{itE_{J}}.$ So for the open
JCQ we have 
\begin{align}
\sigma _{10}\left( t\right) & =\frac{1}{2}\rho _{10}\left( 1-e^{-B^{2}\left(
t\right) }\right) \left( 1-e^{itE_{J}}\right) ,  \notag \\
\sigma _{11}\left( t\right) & =\frac{1}{2}\left( 1-e^{-B^{2}\left( t\right)
}\right) \left( \rho _{00}-\rho _{11}\right) .  \label{eq30}
\end{align}%
Thus, we have%
\begin{align}
\left\| \sigma \left( t\right) \right\| _{\lambda }& =\frac{1}{2}\left(
1-e^{-B^{2}\left( t\right) }\right)  \notag \\
& \left\{ \left( \rho _{00}-\rho _{11}\right) ^{2}+4\left| \rho _{10}\right|
^{2}\sin ^{2}\frac{E_{J}t}{2}\right\} ^{\frac{1}{2}}.  \label{eq31}
\end{align}%
From Eq.(\ref{eq31}) we know that a pure state 
\begin{equation}
\rho _{1}\left( 0\right) =\left[ 
\begin{tabular}{ll}
1 & 0 \\ 
0 & 0%
\end{tabular}%
\right] \text{ or }\rho _{2}\left( 0\right) =\left[ 
\begin{tabular}{ll}
0 & 0 \\ 
0 & 1%
\end{tabular}%
\right]  \label{eq32}
\end{equation}%
will make the $\left\| \sigma \left( t\right) \right\| _{\lambda }$ into $%
D\left( t\right) $. In the following calculations we use the initial state $%
\rho _{1}\left( 0\right) $ and choose the Josephson energy $E_{J}=51.8$ $\mu 
$eV according to Ref. \cite{charge-qubit}.%
\begin{eqnarray*}
&& \\
&&Fig.1 \\
&&
\end{eqnarray*}%
\begin{eqnarray*}
&& \\
&&Fig.2 \\
&&
\end{eqnarray*}%
\begin{eqnarray*}
&& \\
&&Fig.3 \\
&&
\end{eqnarray*}%
By using Eqs. (\ref{eq11},\ref{eq15},\ref{eq18}), (\ref{eq25}), and (\ref%
{eq31}) we can plot the decoherence of JCQ versus time $t$. In Fig.1 we plot
the decoherence from Ohmic noise of the JCQ versus time $t$. This
decoherence is mainly derived from the higher frequency parts of the Ohmic
noise. Here, we confine the frequencies of bath modes in $1$ GHz$\sim 50$
GHz this is a very broad band of frequencies of the environment noise. The
lowest line in Fig.1 plots the decoherence at temperature $T_{1}=0.0300$ K.
It is shown that when $t<0.1,$ namely, $\tau <6.582\times 10^{-11}$ s the
decoherence is much smaller than DiVincenzo criterion $10^{-4}$. The middle
and the upmost lines plot the decoherence at temperature $T_{2}=0.1500$ K
and $T_{3}=0.1875$ K. From the plots we see that the decoherence will
increase with the increasing of the temperature $T$ but it cannot increase
to the maximum value of the DiVincenzo criterion permitting at $T=0.1875$ K
within $t=0.02$ ( $\tau \approx 12.7$ ps). Fig.2 shows the decoherence from $%
1/f$ noise of the JCQ in different $\alpha _{f}.$ It is shown that this
decoherence mainly derived from the low frequency parts of the $1/f$ noise
and it is sensitive to the changing of $\alpha _{f}$. In the calculations,
we take the frequencies of the environment modes from $1$ kHz to $1$ GHz the
numerical simulation shows that the spectrum is broad enough$.$ In the Fig.2
the lowest, middle, and upmost lines plot the decoherence at $\alpha
_{f}=1.0\times 10^{-7},$ $1.1\times 10^{-7}$ and $1.2\times 10^{-7}.$ A
further calculation shows that if only $\alpha _{f}>5.0\times 10^{-8}$ the
decoherence will affect the JCQ as the qubit for making quantum computers
because the decoherence is bigger than the DiVincenzo low decoherence
criterion in the single operation time. So we should decrease the $\alpha
_{f}$ in order to make JCQ be a qubit for quantum computation. In Fig.3 we
simulate three kinds of cases where both the Ohmic and $1/f$ noise are
considered. Here, we set $T=30$ mK and $\alpha _{f}=5\times 10^{-8}$ (the
lowest line), $\alpha _{f}=4\times 10^{-8},$ (the middle line) and $\alpha
_{f}=3\times 10^{-8}$ (the upmost line). The frequencies of bath modes is
set in $1$ kHz$\sim 1$ GHz for $1/f$ noise and in $1$ GHz$\sim 50$ GHz$\ $%
for Ohmic noise. It is shown that when the $\alpha _{f}$ do not exceed $%
5\times 10^{-8}$, within $\tau =12.7$ ps the decoherence from the Ohmic
noise and $1/f$ noise is endurable.

\section{Conclusions}

In this paper we investigated the short-time decoherence results from the
Ohmic and $1/f$ noise of the JCQ. It is shown that the decoherence from $1/f$
noise is larger than that from Ohmic noise to the JCQ model. To the Ohmic
noise the higher frequency parts play a major role to the decoherence. We
take a larger range of the frequencies of the environment modes in our
numerical simulation than usually proposed. It is shown that in usually
experimental temperature this decoherence is not serious comparing to the
DiVincenzo criterion. Unlike the Ohmic noise case, to the $1/f$ noise the
decoherence is mainly determined by the lower frequency parts. It is shown
that when $\alpha _{f}\lesssim 5\times 10^{-8},$ within $\tau =12.7$ ps this
decoherence is also not serious according to the DiVincenzo criterion.
However, the present experiments show that the value of $\alpha _{f}$ is
between $10^{-7}$ and $10^{-6}.$ Thus, if we wish to make the JCQ become a
optimal qubit model we should decrease the value of $\alpha _{f}$. The value
of $\alpha _{f}$ is related to the temperature. It may decrease with the
decreasing of temperature. Carefully finding out the correlation of the
value of $\alpha _{f}$ and other parameters (include the temperature) is a
important task for our investigating the decoherence of JCQ model in the
future.

\begin{acknowledgement}
The project was supported by National Natural Science Foundation of China
(Grant No. 10347133) and Ningbo Youth Foundation (Grant No. 2004A620003).
\end{acknowledgement}

\section{Figs. captions}

Fig.1 The decoherence from Ohmic noise within short time. The lowest, middle
and upmost lines correspond to $T_{2}=0.0300$ K$,$ $T_{2}=0.1500$ K and $%
T_{3}=0.1875$ K. Here, the frequencies of bath modes is set in $1$ GHz$\sim
50$ GHz$.$

Fig.2 The decoherence from $1/f$ noise within short time. The lowest, middle
and upmost lines correspond to $\alpha _{f}=1.0\times 10^{-7},$ $\alpha
_{f}=1.1\times 10^{-7}$, and $\alpha _{f}=1.3\times 10^{-7}.$ Here, the
frequencies of bath modes is set in $1$ kHz$\sim 1$ GHz$.$

Fig.3 The decoherence from Ohmic noise and $1/f$ noise within short time.
The lowest, middle and upmost lines correspond to $\alpha _{f}=3\times
10^{-8},$ $\alpha _{f}=4\times 10^{-8}$, and $\alpha _{f}=5\times 10^{-8}.$
Here, the temperature is set $T=30$ mK, the frequencies of bath modes is set
in $1$ kHz$\sim 1$ GHz for $1/f$ noise and in $1$ GHz$\sim 50$ GHz$\ $for
Ohmic noise.


\begin{thebibliography}{99}
\bibitem{note} In Refs.\cite{open-JCQ-model,Liang02}, the express $%
R_{Q}=\left( 2e\right) ^{2}/h$ is incorrect, but the value $R_{Q}\approx 6.5$
k$\Omega $ is correct.
\end{thebibliography}
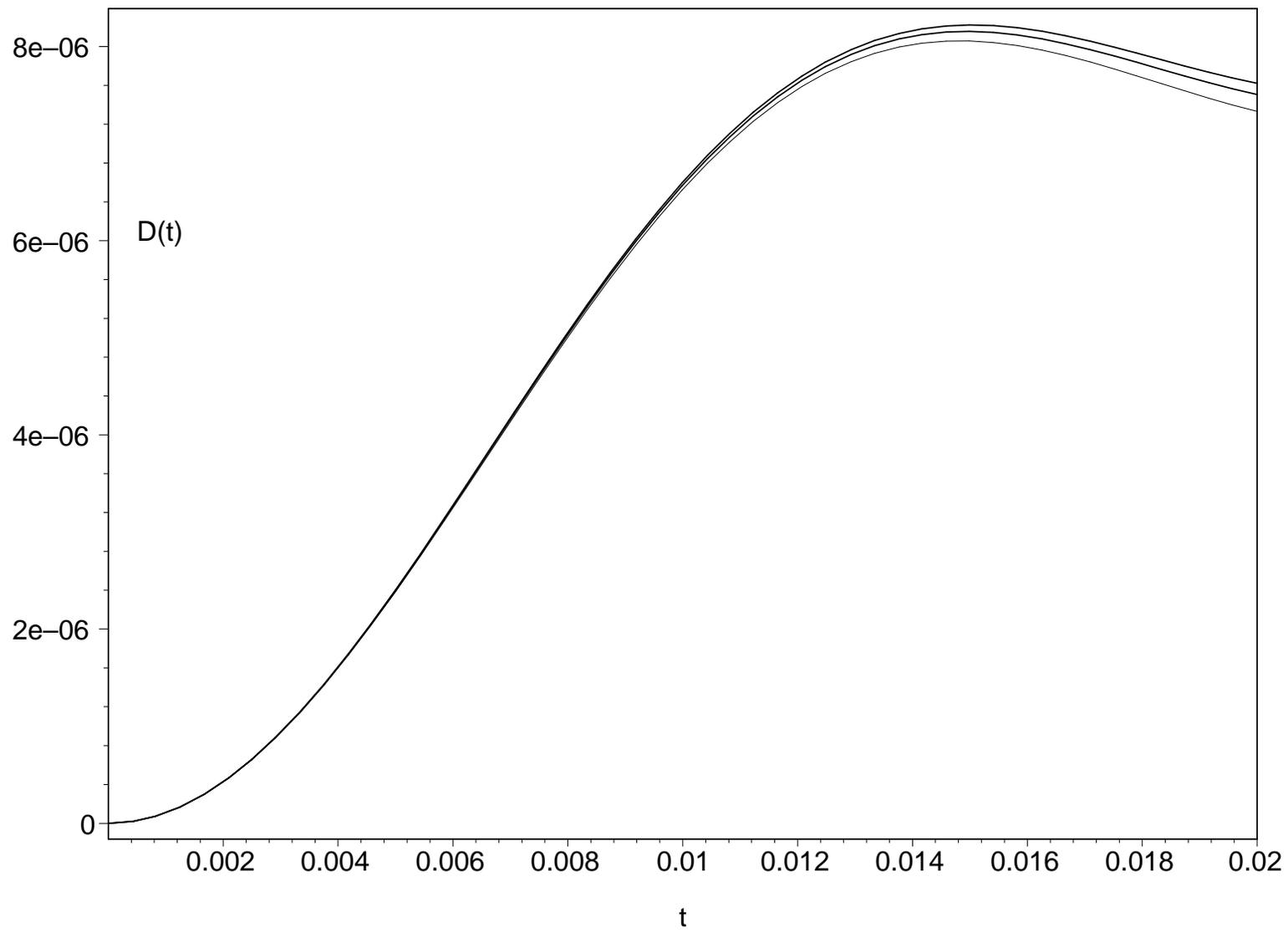
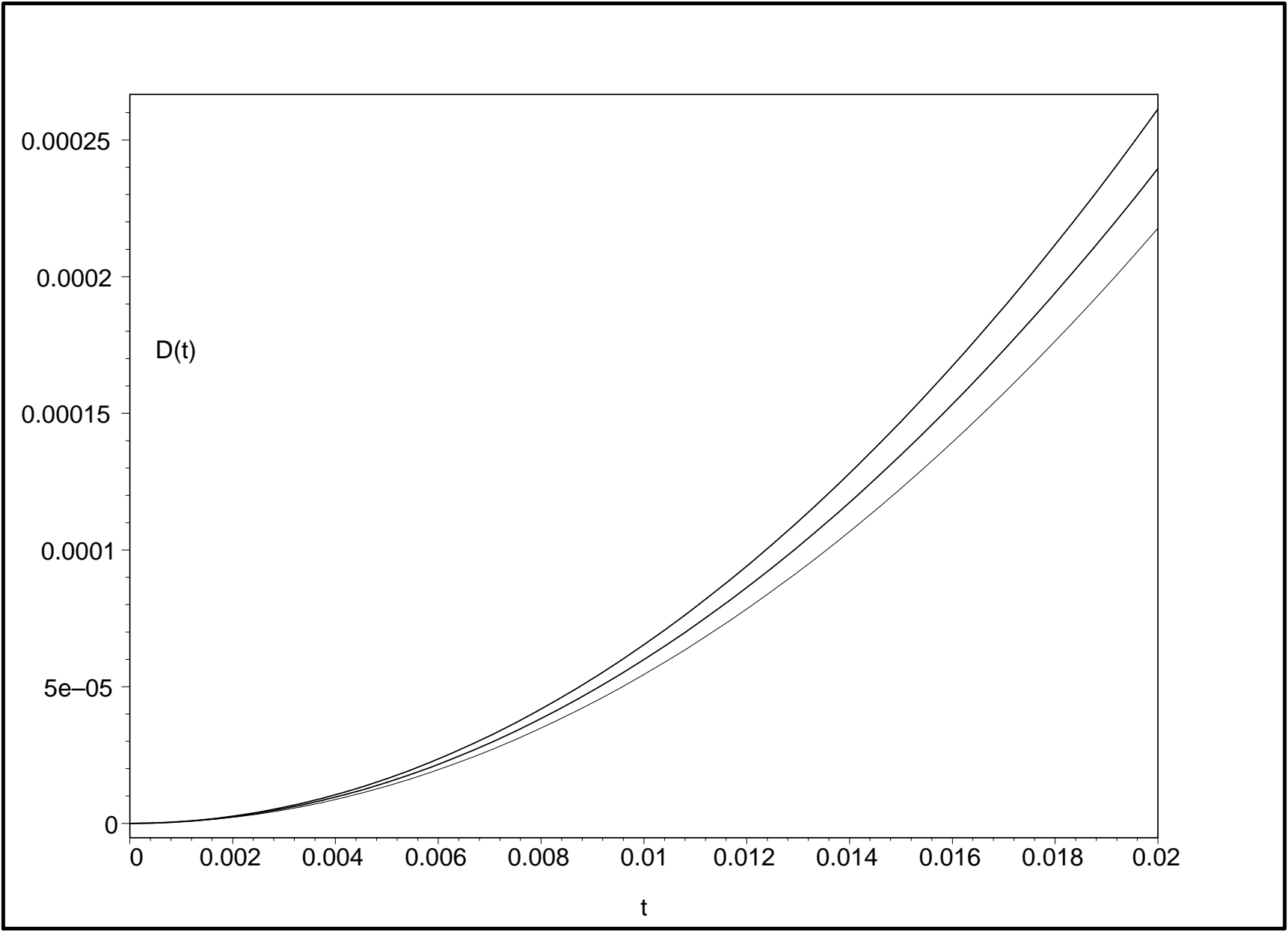
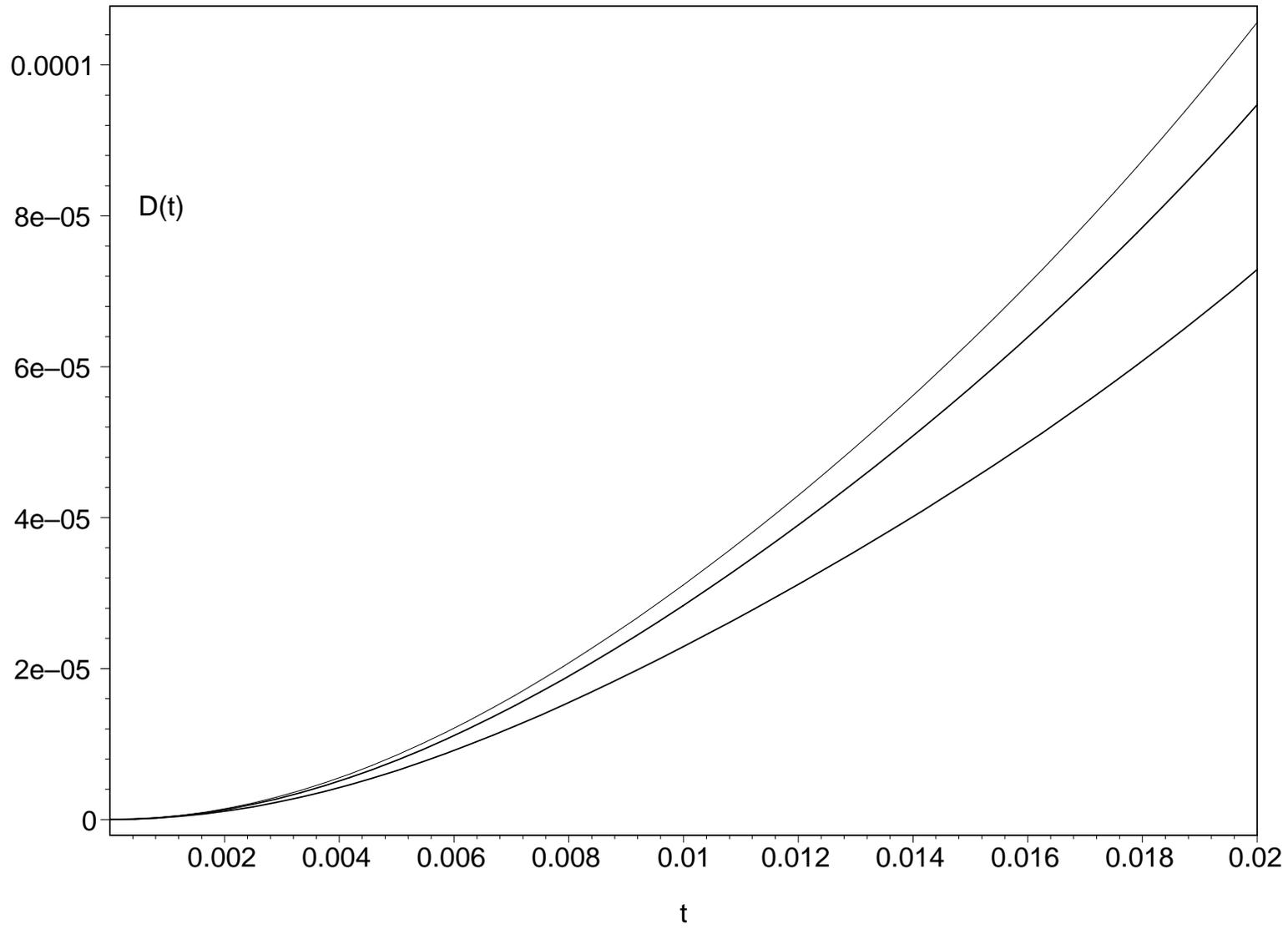
\end{document}